\documentclass[twocolumn,epjc3]{svjour3mod}  

\RequirePackage[T1]{fontenc}
\RequirePackage{graphicx}
\RequirePackage{mathptmx}      
\RequirePackage{latexsym}
\RequirePackage[colorlinks,citecolor=blue,urlcolor=blue,linkcolor=blue]{hyperref}
\RequirePackage{amsmath,amssymb}
\RequirePackage{ulem}
\RequirePackage{booktabs}
\RequirePackage{morefloats}

\newcommand{\diff}{\text{d}}
\newcommand{\Lagr}{\mathcal{L}}
\newcommand{\Order}{\mathcal{O}}
\newcommand{\V}{\mathcal{V}}
\newcommand{\spp}{s}
\newcommand{\GeV}{\,\text{GeV}}
\newcommand{\MeV}{\,\text{MeV}}
\newcommand{\keV}{\,\text{keV}}
\newcommand{\fm}{\,\text{fm}}
\newcommand{\dof}{\text{dof}}
\newcommand{\Br}{\mathcal{B}}
\newcommand{\bsp}{\begin{sloppypar}}
\newcommand{\esp}{\end{sloppypar}}
\newcommand{\nl}{\notag\\}
\renewcommand{\arraystretch}{1.12}

\begin{document}

\title{The branching ratio $\omega\to \pi^+\pi^-$ revisited}

\author{
C.~Hanhart\thanksref{addrJ}
\and
S.~Holz\thanksref{addrBN1}
\and
B.~Kubis\thanksref{addrBN1,addrBN2,e3} 
\and
A.~Kup{\'s}{\'c}\thanksref{addrK1,addrK2}
\and
A.~Wirzba\thanksref{addrJ,addKavli}
\and
C.~W.~Xiao\thanksref{addrJ}
}

\thankstext{e3}{e-mail: kubis@hiskp.uni-bonn.de}

\institute{Institut  f\"{u}r Kernphysik, Institute for Advanced Simulation,
           and J\"ulich Center for Hadron Physics, Forschungszentrum J\"ulich,  
           52425 J\"{u}lich, Germany  \label{addrJ}
           \and
           Helmholtz-Institut f\"ur Strahlen- und Kernphysik, 
           Universit\"at Bonn, 53115  Bonn, Germany \label{addrBN1}
           \and
           Bethe Center for Theoretical Physics,
           Universit\"at Bonn, 53115  Bonn, Germany \label{addrBN2}
           \and
           Division of Nuclear Physics, Department of Physics and Astronomy,
           Uppsala University,  Box 516, 75120 Uppsala, Sweden  \label{addrK1}
           \and
           High Energy Physics Department,
           National Centre for Nuclear Research, 
           ul.\ Hoza 69, 00-681 Warsaw, Poland  \label{addrK2}
           \and
           Kavli Institute for Theoretical Physics, Kohn Hall, University of California,
           Santa Barbara, CA 93106-4030, USA \label{addKavli}
}

\date{}

\maketitle

\begin{abstract}\bsp
We analyze the most recent data for the pion vector form factor in the timelike region,
employing a model-independent approach based on dispersion theory. 
We confirm earlier observations about the inconsistency of different modern high-precision data sets.
Excluding the BaBar data, we find an updated
value for the isospin-violating branching ratio $\Br(\omega\to\pi^+\pi^-) = (1.46\pm 0.08) \times 10^{-2}$.
As a side result, we also extract an improved value for the pion vector or charge radius, 
$\sqrt{\langle r_V^2\rangle} = 0.6603(5)(4)\fm$,
where the first uncertainty is statistical as derived from the fit, while the second estimates
the possible size of nonuniversal radiative corrections.
In addition,
we demonstrate that modern high-quality data for the decay $\eta'\to \pi^+\pi^-\gamma$ 
will allow for an even improved determination of the transition strength $\omega\to\pi^+\pi^-$. 
\esp

\medskip
\noindent\textbf{An erratum is appended at the end of this article.}

\end{abstract}

\section{Introduction}

In recent years interest in high-quality pion form factor data below $\spp=1\GeV^2$ has increased tremendously, since 
it provides a crucial input to quantify the standard model prediction for the hadronic contribution to the 
muon anomalous magnetic moment  (see, e.g., Refs.~\cite{Miller:2012opa,Blum:2013xva,Benayoun:2014tra} and references therein) 
and the dispersion integral that
needs be evaluated in this context puts a lot of weight on the low-energy transition $\gamma^*\to$ hadrons. 
To make the most of the existing data, it is compulsory to employ model-independent theoretical tools
that allow for an appropriate parametrization of the data, but also for a judgment on their consistency.
For the two-pion contributions to the above-mentioned transitions the appropriate tool is again dispersion theory, for
it allows one to use the high-quality information available for pion--pion 
scattering~\cite{Ananthanarayan:2000ht,Colangelo:2001df,GarciaMartin:2011cn,Caprini:2011ky} 
in the form factor analysis in a way consistent with analyticity and unitarity.
The strong impact these theoretical constraints can have on our understanding of the pion form factor has been emphasized
and used to good effect several times 
before~\cite{DeTroconiz:2001rip,deTroconiz:2004yzs,Leutwyler:2002hm,Colangelo:2003yw,Ananthanarayan:2013zua,Ananthanarayan:2016mns,Hoferichter:2016duk},
with some of those references very close in spirit to what we are attempting here.

\bsp
We exemplify the power of this formalism by an analysis of the most recent data sets for the pion vector form factor extracted from 
measurements of $e^+e^-\to \pi^+\pi^-$, with the specific goal to extract an update on the partial width for $\omega\to \pi^+\pi^-$. 
As a side result we also determine an updated value for the pion vector or charge radius. 
Since final-state interactions are universal within the same scheme, we also
propose to analyze the reaction $\eta'\to \pi^+\pi^-\gamma$: not only will high-quality data for this reaction become available
from different experiments in the very near future, but also it is shown to provide additional, independent access to 
the $\omega\to\pi^+\pi^-$ transition strength. To illustrate the potential accuracy of such a determination once the new data are available,
we here analyze pseudo-data generated according to preliminary results from BESIII~\cite{Fang:2015vva}.
\esp

One key feature of the formalism employed here is that it makes maximal use of the universal phase introduced by the pion--pion
final-state interactions. In particular, we do not have the freedom to add Breit--Wigner functions with arbitrary relative
phases. This allows us to extract the relevant amplitudes in a controlled fashion.  
As a side note, we illustrate the reaction-dependence of Breit--Wigner functions explicitly by demonstrating
that a (constant) complex phase in the coupling and 
a shift of the $\omega$ mass parameter  lead to similar effects on the observables.  

This paper is organized as follows. 
In Sect.~\ref{sec:formalism}, we lay out the necessary formalism, introducing the dispersive representations
of both the pion vector form factor and the $\eta'\to\pi^+\pi^-\gamma$ decay amplitude and showing how the 
parameters of the $\rho$--$\omega$ mixing signals can be related to the decay width of $\omega\to\pi^+\pi^-$.
This is followed in Sect.~\ref{sec:results} by a detailed discussion
of the results for the $\omega\to\pi^+\pi^-$ branching fraction,
obtained from elaborate fits both to various $e^+e^-\to\pi^+\pi^-$ data sets as well as the BESIII pseudo-data for the decay $\eta'\to \pi^+\pi^-\gamma$.
Section~\ref{sec:radius} presents our findings on the pion vector radius.  
We close with a summary.

\section{Formalism} \label{sec:formalism}

\subsection{Matrix elements}

The pion vector form factor $F_V(\spp)$, which describes the reaction $e^+e^-\to\pi^+\pi^-$, is  
defined by the vector current matrix element 
\begin{align}
\langle \pi^+(k_+)\pi^-(k_-)|\V^\mu|0\rangle &=  e(k_+- k_-)^\mu F_V(\spp), \nl
 \spp &= q^2, \quad q^\mu  =k_+^\mu +k_-^\mu,  \label{eq:def_FV}
\end{align}
where  $e>0$ is the unit of the electric charge. 
Throughout this work
we apply the definition $\V^\mu = -{ \delta {\cal L}_{\text{int}}}/{\delta A_\mu}$, with the photon field $A_\mu$.
For the pion fields, we use the Condon--Shortley sign convention 
$\pi^\pm = \mp (\pi^1\mp i\pi^2)/\sqrt{2}$. 

The matrix element describing the decay $\eta'\to\pi^+\pi^-\gamma$ 
in the $P$-wave approximation can be written as\footnote{We 
use the sign and phase assignments according to  Refs.~\cite{Har_Yam_2003,Klingl:1996by},
adapted for the fact that both works implicitly
assume a negative value for $e$ and do not follow the Condon--Shortley convention.}
\begin{equation}
\langle \pi^+(k_+)\pi^-(k_-)|\V_\mu|\eta'(p)\rangle =  \epsilon_{\mu\nu\alpha\beta}  p^\nu  k_+^\alpha \,   k_-^\beta \,
f_1(\spp)  \label{eq:Fetapipigamma} 
\end{equation}
(see Ref.~\cite{Kubis:2015sga} for a definition of the partial-wave expansion).
We define $\epsilon_{\mu\nu\alpha\beta}$ such that $\epsilon_{0123} = +1$.
The corresponding differential decay rate is given by
\begin{equation}
\frac{\diff\Gamma(\eta'\to\pi^+\pi^-\gamma) }{\diff \sqrt{\spp}}  = \left | f_1(\spp) \right|^2 \Gamma_1(\spp) ,
\end{equation}
where the function 
\begin{equation}
\Gamma_1(\spp) = \frac{4}{3}\bigg( \frac{m_{\eta'}^2 - \spp}{16\pi m_{\eta'}}
\sqrt{\spp- 4 m_\pi^2}\bigg)^3
\end{equation}
collects the phase-space terms and kinematical factors of the modulus squared of the invariant matrix element for the 
point-particle case~\cite{Stollenwerk:2011zz}, with $m_{\eta'}$ and $m_\pi$ denoting the mass of the $\eta'$ and the pion, respectively.

\subsection{Universality of final-state interactions and dispersive representations} \label{sec:FSI}

\bsp
We base our analysis on the fact that as a result of unitarity, all elastic pion--pion ($\pi\pi$) 
interactions of a definite partial wave are largely determined by a single, universal
function given in terms of the corresponding $\pi\pi$ phase shift---the Omn\`es function $\Omega(\spp)$,
depending only on $\spp$, the squared invariant mass of the outgoing pion pair.  
For pion pairs with relative angular momentum $L=1$, it is given by
\begin{equation}
\Omega(\spp) = 
\exp\bigg\{\frac{\spp}{\pi}\int_{4m_\pi^2}^\infty\diff x\frac{\delta_1(x)}{x(x-\spp {-i\epsilon})}\bigg\},
\end{equation}
where $\delta_1 (\spp)$ denotes the pion--pion $P$-wave phase shift.
The Omn\`es function captures the physics of the $\rho$-meson, encoded in the phase shift in a model-independent way, 
thus eschewing the need to use a model like vector-meson dominance.
Recent phase-shift analyses based on sophisticated dispersive analyses are available from the Madrid~\cite{GarciaMartin:2011cn} 
and Bern~\cite{Caprini:2011ky} groups
in an energy range from threshold up to about $1.4\GeV$.
In our analysis, we continue these phase shifts smoothly to an asymptotic value of $\pi$ above $1.3$
and $1.42\GeV$, respectively,
in order to fix $\Omega(\spp)$ completely. 
As we are interested in an evaluation of the Omn\`es function only for energies below $1\GeV$, 
the precise rate at which this limiting value is approached is immaterial:
it leads to changes in the Omn\`es function that can be absorbed
in the parametrizations used in this work.

In  Refs.~\cite{Stollenwerk:2011zz,Hanhart:2013vba},
the universality of the final-state interactions was used to express $F_V(\spp)$ and $f_1(\spp)$ in the forms
\begin{equation}
F_V(\spp) = R(\spp)\Omega(\spp) , \qquad 
f_1(\spp) = P(\spp)\Omega(\spp) . \label{eq:def_R+P}
\end{equation}
The functions $R(\spp)$ and $P(\spp)$ must be real and free of right-hand cuts in the elastic region; 
in Refs.~\cite{Stollenwerk:2011zz,Hanhart:2013vba} they were assumed to be linear polynomials,
which was demonstrated to be sufficient for the (isospin-related) vector form factor featuring in 
the decay $\tau^-\to\pi^-\pi^-\nu_\tau$, as well as the decay $\eta\to\pi^+\pi^-\gamma$~\cite{Adlarson:2011xb,Babusci:2012ft} 
similar to the $\eta'$ transition.
The universal phase that $F_V(\spp)$ and $f_1(\spp)$ share with the Omn\`es function, given by $\delta_1(\spp)$,
is a consequence of Watson's final-state theorem~\cite{Watson:1954uc}.
The formalism for the $\eta'$ decay was improved further in Ref.~\cite{Kubis:2015sga}, 
where it was shown that $P(\spp)$ contains a left-hand cut induced by
tensor-meson ($a_2(1320)$) exchange, which in the physical decay region can be approximated to very good precision by
the inclusion of a quadratic term in $P(\spp)$.

However, the expressions given so far
ignore the contribution from the $\omega$-meson, which can also decay into the $\pi^+\pi^-$ final state via 
isospin-violating interactions. 
While we assume isospin symmetry everywhere else, this particular isospin-breaking effect is enhanced by a small
energy denominator, as the $\omega$-resonance is very narrow and close in mass to the $\rho$, the dominant resonant 
enhancement of the $\pi\pi$ $P$-wave amplitude.
It is well-known that the inclusion of this mechanism, often named $\rho$--$\omega$ mixing 
(see also Refs.~\cite{Urech:1995ry,Kucukarslan:2006wk} for effective field theory approaches to this phenomenon), 
is essential for an accurate description of the vector form factor in $e^+e^-\to\pi^+\pi^-$.
In this paper we extend the formalism of Refs.~\cite{Kubis:2015sga,Stollenwerk:2011zz,Hanhart:2013vba}
to include this effect, which gives access to the $\omega\to\pi^+\pi^-$ transition strength.
\begin{figure} 
 \centering
   \includegraphics*[width=\linewidth]{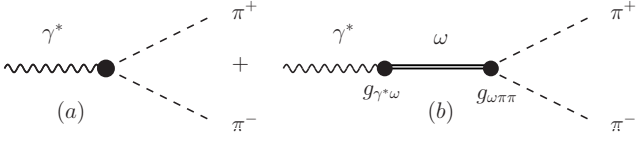}
   \caption{Diagrammatic representation of the reaction $\gamma^* \to \pi^+ \pi^-$.
      The pions from both diagrams undergo final-state interactions that are not shown explicitly.    
      }\label{fig:diapi}
\end{figure}
\begin{figure} 
 \centering
   \includegraphics*[width=\linewidth]{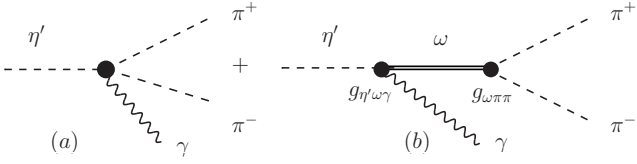}
   \caption{Diagrams contributing to the reaction $\eta'\to\pi^+\pi^-\gamma$.
   The pions from both diagrams undergo final-state interactions that are not shown explicitly.        
   }\label{fig:diaeta}
\end{figure}
The contributions of the $\omega$ are shown diagrammatically in Figs.~\ref{fig:diapi} and \ref{fig:diaeta}.
In both cases the outgoing pion pair undergoes final-state interactions in the $P$-wave,
which are universal and controlled by the Omn\`es function. In the $\omega$-channel
on the other hand, the use of a Breit--Wigner function appears to be justified,
since the $\omega$ total width is small, $\Gamma_{\omega}^{\rm tot} = (8.49\pm0.08)\MeV$~\cite{Olive:2016xmw}. 
Our generalization of the polynomials $R(\spp)$ and $P(\spp)$ now reads
\begin{align}
R(\spp) &=1+\alpha_V\, \spp + \frac{\kappa_1 \, \spp}{m_\omega^2-\spp-im_\omega \Gamma_{\omega}^{\rm tot}} , \label{eq:rspp} \\
P(\spp) &= A \big(1 + \alpha\, \spp + \beta\, \spp^2\big) + \frac{\kappa_2}{m_\omega^2-\spp-im_\omega \Gamma_{\omega}^{\rm tot}}  , \label{eq:ppspp}
\end{align}
where $m_\omega$ denotes the $\omega$ mass and
$\alpha_V$, $\alpha$, $\beta$, $\kappa_1$, and $\kappa_2$ are constants to be determined from a fit to data.
Equations~\eqref{eq:rspp} and \eqref{eq:ppspp} are correct to leading order in isospin violation. 
Unitarity dictates that after the transition
from the $\omega$-meson to a pion pair, the phase induced by the final-state
interaction must again be equal to that of pion--pion $P$-wave scattering.
This leads to the requirement that both $\kappa_1$ and $\kappa_2$ are real-valued.
This statement also holds up to higher orders in isospin violation, 
which are expected to provide negligible corrections.
Similar ans\"atze for the $\rho$--$\omega$ mixing term were used 
before frequently~\cite{Leutwyler:2002hm,Ananthanarayan:2013zua,Ananthanarayan:2016mns,Gardner:1997ie,Hanhart:2012wi,Daub:2015xja};
we employ the sign convention of Ref.~\cite{Daub:2015xja}. 

We checked that the results extracted from the pion vector form factor
using Eqs.~\eqref{eq:def_R+P} and \eqref{eq:rspp} are stable against the inclusion
of an additional term quadratic in $\spp$ in $R(\spp)$, as we will discuss in detail in Sect.~\ref{sec:results}. 
A detailed study of the effect of an $s^3$-term on an analysis of the $\eta'$-decay using Eq.~\eqref{eq:ppspp} 
is postponed until a final data set becomes available; the effect on the extraction of $\kappa_2$ is expected 
to be negligible.

\subsection{The relation to $\Gamma(\omega\to\pi^+\pi^-)$}
 
The parameters $\kappa_{1/2}$ are proportional to the coupling
strength of the $\omega$ to two pions $g_{\omega \pi\pi}$, see Figs.~\ref{fig:diapi} and \ref{fig:diaeta}, 
which in turn can be related
to the partial decay width $\omega\to \pi^+\pi^-$. To calculate the factor 
of proportionality we need to utilize proper vertex functions as outlined below.
This subsection is devoted to establishing the connection in such a way that 
$\Gamma(\omega\to\pi^+\pi^-)$ can be determined from an extraction of $\kappa_{1/2}$
in a fit to the available data.

In order to connect $g_{\omega \pi\pi}$ to the quantities $\kappa_1$ and $\kappa_2$ defined in Eqs.~\eqref{eq:rspp} and \eqref{eq:ppspp}, we first derive the $\omega\to \pi^+\pi^-$ vertex
\begin{equation}
  \langle \pi^+(k_+) \pi^-(k_-) |\Lagr_{\omega\pi\pi}| \omega (q) \rangle = -g_{\omega\pi\pi} \, \epsilon_{\mu}^{(\omega)}(q) (k_+- k_-)^\mu  ,
  \label{eq:omepip}
\end{equation}
with $\epsilon^{(\omega)}_\mu(q)$ the pertinent polarization vector of the vector meson  $\omega$ of
momentum $q$, 
from the interaction Lagrangian
\begin{equation}
  \Lagr_{\omega\pi\pi} = i g_{\omega\pi\pi} \left(\pi^- \partial^\mu\pi^+ - \pi^+\partial^\mu \pi^-\right)\omega_\mu ,
\end{equation}
which is the analog of the $\gamma\pi\pi$ Lagrangian with $g_{\omega\pi\pi}$ 
and the vector field $\omega_\mu$ of the $\omega$-meson 
taking over the role of the charge $e$ and the photon field $A_\mu$, respectively; see also Ref.~\cite{Klingl:1996by}, as well as Ref.~\cite{Meissner:1987ge} for a comprehensive overview of vector-meson Lagrangians.  
Furthermore, we need the coupling of the $\omega$ to a virtual photon as well as the vertex for $\eta'\to \omega\gamma$.
For the former we use the  effective Lagrangian~\cite{Klingl:1996by,O'Connell:1995xx}
\begin{equation}
  \Lagr_{\omega\gamma} = - \frac{e}{2} g_{\omega\gamma} F^{\mu\nu}\omega_{\mu \nu} ,
\end{equation}
where $F_{\mu\nu}=\partial_\mu A_\nu - \partial_\nu A_\mu$ is the electromagnetic field strength tensor 
and $\omega_{\mu\nu}= \partial_\mu \omega_\nu - \partial_\nu \omega_\mu$.
Contrary to standard vector-meson dominance formulations, we couple the $\omega$-meson (with the
coupling strength $g_{\omega\gamma}$)
to the electromagnetic field strength tensor 
and not to the vector field in order to ensure gauge invariance directly on the level of the vertex. The additional derivatives that accompany
this choice are the origin of the factor $\spp$  ($=q^2$) in Eq.~\eqref{eq:rspp}; the corresponding vertex reads
\begin{equation}
  \langle \omega (q)|\V_\mu|0\rangle = e g_{\omega\gamma}\, q^2   \epsilon^{(\omega)}_\mu(q)  .
  \label{eq:vgamq}
\end{equation}
Furthermore, when taking $q^2 = m_\omega^2$ and neglecting the electron mass, the $e^+e^-$
decay width of the $\omega$ is given by 
\begin{equation}
  \Gamma (\omega \to e^+ e^-) = \frac{4\pi \alpha_{\rm em}^2}{3} g_{\omega\gamma}^2 m_\omega  , 
\end{equation}
with $\alpha_{\rm em} \approx 1/137.036$ the electromagnetic fine structure constant.
Using  $\Gamma (\omega \to e^+ e^-) = (0.60 \pm 0.02)\keV$~\cite{Olive:2016xmw}, we obtain
\begin{equation}
|g_{\omega\gamma}| = (5.9\pm 0.1)\times 10^{-2}  . \label{eq:g_omegagamma}
\end{equation}
In this way one finds for the transition $\gamma^*\to\omega\to\pi^+\pi^-$, corresponding to Fig.~\ref{fig:diapi}b, 
\begin{equation}
\left.\langle \pi^+(k_+)\pi^-(k_-)|\V^\mu|0\rangle\right|_{\omega} =  \frac{e \,g_{\omega\pi\pi}\, g_{\omega\gamma}\, \spp}{m_\omega^2-\spp-im_\omega \Gamma_{\omega}^{\rm tot}}(k_+- k_-)^\mu  ,
\end{equation}
where we used $\sum \epsilon_{\mu}^{(\omega)}(q) \epsilon_{\nu}^{(\omega)}(q) = -g_{\mu\nu}+q^\mu q^\nu/m_\omega^2$. 
Comparison with Eqs.~\eqref{eq:def_R+P} and \eqref{eq:rspp} allows us to identify 
\begin{equation}
\kappa_1 =  g_{\omega\pi\pi}\, g_{\omega\gamma} .
\label{kappa1}
\end{equation}
Since we have extracted $g_{\omega\gamma}$ above, we may quantify $g_{\omega\pi\pi}$ once $\kappa_1$ is fixed from a
fit to form factor data.
Note that Eq.~\eqref{eq:g_omegagamma} does not fix the sign of $g_{\omega\gamma}$, which therefore would
also leave the sign of $g_{\omega\pi\pi}$ undetermined.
However, if $\rho$-dominance is used to model the isospin-conserving part of the pion form factor and
the signs of $g_{\omega\gamma}$ and $g_{\rho\gamma}$ are assumed equal as suggested by $SU(3)$ flavor symmetry, 
then positive values for $\kappa_1$ (which we will find empirically in the following section)
show that $g_{\omega\pi\pi}$ has the same sign as
a conventional $g_{\rho\pi\pi}$ coupling. Accordingly we assume $g_{\omega\pi\pi}$ to be positive in our analysis.
Obviously, the observable $\omega\to\pi^+\pi^-$ partial width or branching fraction is independent of this sign.

The expression for the $\eta'\omega \gamma$ vertex, again according to the sign and phase conventions
of Refs.~\cite{Har_Yam_2003,Klingl:1996by}, is
\begin{equation}
  \langle \omega (q)| \V^\mu| \eta' (p)\rangle = g_{\eta'\omega\gamma} \, \epsilon^{\mu\nu\alpha\beta} p_\nu q_\alpha \epsilon_{\beta}^{(\omega)}(q)  .
  \label{eq:etapq}
\end{equation}
The coupling constant $ g_{\eta'\omega\gamma}$ can be determined using 
\begin{equation}
\Gamma(\eta'\to\omega\gamma)=\frac{g_{\eta'\omega\gamma}^2}{32\pi}\bigg( \frac{m_{\eta'}^2-m_\omega^2}{m_{\eta'}} \bigg)^3 ,
\end{equation}
and the measured decay width $\Gamma(\eta'\to\omega\gamma)=(5.17\pm 0.35)\keV$~\cite{Olive:2016xmw},
which leads to
\begin{equation}
g_{\eta'\omega\gamma} = -(0.127 \pm 0.004) \GeV^{-1}  , \label{eq:getapomegagamma}
\end{equation}
where the negative sign is consistent with the speci\-fications in Refs.~\cite{Har_Yam_2003,Klingl:1996by}.
Analogously to the steps followed above, we may combine the vertex given in Eq.~\eqref{eq:etapq} with Eq.~\eqref{eq:omepip} to find
\begin{align}
\langle \pi^+(k_+)&\pi^-(k_-)|\V^\mu|\eta'(p)\rangle\big|_{\omega} \nl
&=  \frac{g_{\eta'\omega\gamma} g_{\omega\pi\pi}}{m_\omega^2-\spp -im_\omega \Gamma_{\omega}^{\rm tot}}
\epsilon^{\mu\nu\alpha\beta} p_\nu q_\alpha  (k_+ - k_-)_\beta .
\end{align}
Thus, the comparison with Eqs.~\eqref{eq:Fetapipigamma} and \eqref{eq:ppspp} yields
\begin{equation}
\kappa_2=-2g_{\omega\pi\pi}g_{\eta'\omega\gamma} .
\label{kappa2}
\end{equation}
We will see later that $\kappa_2$ turns out to be positive empirically, such that Eq.~\eqref{eq:getapomegagamma}
shows consistency with the positive sign for $g_{\omega\pi\pi}$ once more.
With these expressions we are prepared to analyze the data for both the pion vector form factor as well
as the decay $\eta'\to\pi^+\pi^-\gamma$.

It should be clear from the discussions above that $g_{\omega\pi\pi}$ only provides the
strength for a pion pair to be produced in the decay of the $\omega$-meson. This pion pair
subsequently undergoes final-state interactions that are parametrized via the complex-valued Omn\`es function $\Omega(\spp)$,
which leads to a significant enhancement of the $\omega$ transition rate, since  
 $\left|\Omega(m_\omega^2)\right|^2\simeq 30$.
Accordingly, the partial decay width for the transition $\omega\to \pi^+\pi^-$ is given by
\begin{equation}
\Gamma(\omega\to\pi^+\pi^-) = \frac{g_{\omega\pi\pi}^2}{48\pi}\frac{\big(m_\omega^2-4m_\pi^2\big)^{3/2}}{m_\omega^2}\left|\Omega(m_\omega^2)\right|^2  .
\label{gammaompp}
\end{equation}
We checked numerically that the results change only marginally if we take the finite $\omega$
mass distribution into account.
\esp

\section{Extracting the branching ratio $\omega\to\pi^+\pi^-$}\label{sec:results}

\subsection{Pion vector form factor}\label{sec:formfactor}
Recent data for the pion vector form factor is available from 
SND~\cite{Achasov:2006vp},
CMD-2~\cite{Akhmetshin:2006bx},
BaBar~\cite{Aubert:2009ad}, 
KLOE~\cite{Ambrosino:2010bv,Babusci:2012rp}, labeled below as KLOE10 and KLOE12, respectively,  and
BESIII~\cite{Ablikim:2015orh}. Up to now, only the first of these data sets
is included in the averages of the Particle Data Group (PDG) for $\Gamma(\omega\to \pi^+\pi^-)$, 
and none for the pion vector radius.
As the fitting ranges we chose all data of the mentioned sets from the lowest-energy
point up to $\spp=1\GeV^2$---beyond this energy, effects of the excited $\rho$ resonances start to set in
that can no longer be parametrized by a polynomial (see e.g.\ Fig.~1 of Ref.~\cite{Hanhart:2013vba}).
We use the form factor data provided by the experiments without covariance matrices;
these are available only from KLOE
(for BESIII there is an uncertainty in the overall normalization factor).
We have checked for the KLOE data that the inclusion of the covariance matrices does not change the fit results. 
In addition, the errors of the parameters obtained from the fits to the KLOE data
are not affected when the dominating systematic
uncertainty (near $s=m_\omega^2$) due to unfolding of the initial-state-radiation cross sections is included.
For the main results of our study we employ the Omn\`es function derived from the phase shift 
based on the best-fit values quoted by the Madrid analysis~\cite{GarciaMartin:2011cn};
as a cross-check we also performed fits based on the Bern phase shifts~\cite{Caprini:2011ky}.

\bsp
First we only fit the parameters $\alpha_V$ and $\kappa_1$,
keeping the values for the $\omega$ mass and width fixed to the central values provided by the 
PDG~\cite{Olive:2016xmw}, namely $782.65\MeV$ and $8.49\MeV$, respectively.
The fit parameters as well as the values for $\chi^2$ per degree of freedom are given as Fit~1 
in Table~\ref{tab:fitpvff}.
\begin{table*} 
\centering
\caption{Fit results for the pion vector form factor. 
Fit~1: $\omega$ mass and width fixed; Fit~2: $\omega$ mass allowed to float; 
Fits~I and II: as Fits~1 and 2, but employing the Omn\`es function derived from the phase shifts of the Bern analysis~\cite{Caprini:2011ky}. 
Fits~1-$\rho$ and 2-$\rho$: as Fits~1 and 2, but fitting the mass parameter $m_\rho$ in the phase shift $\delta_1(\spp)$ of Ref.~\cite{GarciaMartin:2011cn} as well. 
Fits~1-$\phi$ and 2-$\phi$: as Fits~1 and 2, but with an Orsay phase.
Fixed parameter values are marked by an asterisk ($^*$). 
[h!]The numbers of data points included in the fits are 
45 for SND~\cite{Achasov:2006vp}, 
28 for CMD-2~\cite{Akhmetshin:2006bx}, 
268 for BaBar~\cite{Aubert:2009ad},
75 for KLOE10~\cite{Ambrosino:2010bv}, 
60 for KLOE12~\cite{Babusci:2012rp}, and 
60 for BESIII~\cite{Ablikim:2015orh}. 
}
\label{tab:fitpvff}
\begin{tabular*}{\textwidth}{@{\extracolsep{\fill}}lllll lllll l@{}}
\toprule
fits & data set & $\alpha_V \times 10$ & $\kappa_1 \times 10^3$ & $m_\omega$ & $\phi$ & $m_\rho$ & $\chi^2/\dof$ & $g_{\omega\pi\pi} \times 10^2$ & $\Br(\omega\to\pi^+\pi^-)$ & $\langle r_V^2 \rangle$ \\
 &  & $[\GeV^{-2}]$ &  & $[\MeV]$ & $[{}^\circ]$ & $[\MeV]$ &  &  & $[\%]$ & $[\text{fm}^2]$ \\
\midrule
Fit 1 
 & SND    & $0.91 (2)$ & $1.73 (4) $ & $782.65  \;^*$ & $0 \;^*$ & $773.6\;^*$  &  3.60  & $2.96 (8) $ & $1.45 (8) $ & $0.4377 (4)$ \\
 & CMD-2  & $0.99 (2)$ & $1.60 (5) $ & $782.65  \;^*$ & $0 \;^*$ & $773.6\;^*$  &  1.96  & $2.73 (10)$ & $1.24 (9) $ & $0.4395 (5)$ \\
 & BaBar  & $0.93 (1)$ & $2.25 (4) $ & $782.65  \;^*$ & $0 \;^*$ & $773.6\;^*$  &  1.52  & $3.84 (9) $ & $2.45 (11)$ & $0.4384 (2)$ \\
 & KLOE10 & $0.81 (1)$ & $1.68 (5) $ & $782.65  \;^*$ & $0 \;^*$ & $773.6\;^*$  &  1.06  & $2.86 (9) $ & $1.35 (9) $ & $0.4353 (3)$ \\
 & KLOE12 & $0.83 (1)$ & $1.46 (10)$ & $782.65  \;^*$ & $0 \;^*$ & $773.6\;^*$  &  1.21  & $2.50 (18)$ & $1.03 (15)$ & $0.4357 (3)$ \\
 & BESIII & $0.91 (4)$ & $1.75 (13)$ & $782.65  \;^*$ & $0 \;^*$ & $773.6\;^*$  &  0.82  & $2.98 (23)$ & $1.48 (23)$ & $0.4378 (9)$ \\
\midrule
Fit 2 
 & SND    & $0.96 (2)$ & $1.80 (4) $ & $781.63 (10)$ & $0 \;^*$ & $773.6\;^*$  &  1.21  & $3.07 (8) $ & $1.57 (9) $ & $0.4388 (4)$ \\
 & CMD-2  & $1.04 (2)$ & $1.67 (6) $ & $782.32 (8) $ & $0 \;^*$ & $773.6\;^*$  &  1.45  & $2.85 (11)$ & $1.35 (10)$ & $0.4406 (6)$ \\
 & BaBar  & $0.93 (1)$ & $2.30 (4) $ & $781.72 (10)$ & $0 \;^*$ & $773.6\;^*$  &  1.16  & $3.93 (9) $ & $2.58 (12)$ & $0.4384 (2)$ \\
 & KLOE10 & $0.81 (1)$ & $1.69 (5) $ & $782.87 (14)$ & $0 \;^*$ & $773.6\;^*$  &  1.05  & $2.88 (10)$ & $1.37 (9) $ & $0.4353 (3)$ \\
 & KLOE12 & $0.83 (1)$ & $1.45 (10)$ & $782.29 (43)$ & $0 \;^*$ & $773.6\;^*$  &  1.21  & $2.48 (18)$ & $1.02 (15)$ & $0.4357 (3)$ \\
 & BESIII & $0.91 (4)$ & $1.78 (13)$ & $781.87 (45)$ & $0 \;^*$ & $773.6\;^*$  &  0.78  & $3.03 (23)$ & $1.53 (23)$ & $0.4378 (9)$ \\
\midrule
Fit I
 & SND    & $0.47 (2)$ & $1.75 (4) $  & $782.65  \;^*$ & $0 \;^*$ & ---  &  4.15  & $2.99 (8) $ & $1.56 (8) $  & $0.4334 (4)$ \\
 & CMD-2  & $0.53 (2)$ & $1.61 (5) $  & $782.65  \;^*$ & $0 \;^*$ & ---  &  1.80  & $2.75 (10)$ & $1.32 (9) $  & $0.4348 (5)$ \\
 & BaBar  & $0.51 (1)$ & $2.38 (3) $  & $782.65  \;^*$ & $0 \;^*$ & ---  &  2.74  & $4.06 (9) $ & $2.89 (13)$  & $0.4346 (2)$ \\
 & KLOE10 & $0.36 (1)$ & $1.83 (5) $  & $782.65  \;^*$ & $0 \;^*$ & ---  &  1.65  & $3.11 (10)$ & $1.69 (11)$  & $0.4308 (2)$ \\
 & KLOE12 & $0.42 (1)$ & $1.68 (10)$  & $782.65  \;^*$ & $0 \;^*$ & ---  &  1.28  & $2.87 (17)$ & $1.44 (17)$  & $0.4321 (3)$ \\
 & BESIII & $0.50 (4)$ & $1.88 (13)$  & $782.65  \;^*$ & $0 \;^*$ & ---  &  0.79  & $3.21 (22)$ & $1.81 (25)$  & $0.4342 (8)$ \\
\midrule
Fit II 
 & SND    & $0.52 (2)$ & $1.82 (4) $  & $781.58 (10)$ & $0 \;^*$ & ---  &  1.41  & $3.11 (8) $ & $1.70 (9) $  & $0.4346 (4)$ \\
 & CMD-2  & $0.59 (2)$ & $1.69 (5) $  & $782.29 (8) $ & $0 \;^*$ & ---  &  1.13  & $2.88 (10)$ & $1.45 (10)$  & $0.4361 (6)$ \\
 & BaBar  & $0.52 (1)$ & $2.44 (3) $  & $781.64 (9) $ & $0 \;^*$ & ---  &  2.29  & $4.17 (9) $ & $3.05 (13)$  & $0.4347 (2)$ \\
 & KLOE10 & $0.36 (1)$ & $1.84 (5) $  & $782.92 (13)$ & $0 \;^*$ & ---  &  1.61  & $3.14 (10)$ & $1.72 (11)$  & $0.4308 (2)$ \\
 & KLOE12 & $0.41 (1)$ & $1.68 (10)$  & $782.46 (33)$ & $0 \;^*$ & ---  &  1.30  & $2.86 (17)$ & $1.43 (17)$  & $0.4321 (3)$ \\
 & BESIII & $0.50 (4)$ & $1.92 (13)$  & $781.80 (42)$ & $0 \;^*$ & ---  &  0.73  & $3.27 (22)$ & $1.88 (26)$  & $0.4342 (8)$ \\
\midrule
Fit 1-$\rho$ 
 & SND    & $0.91 (2)$ & $1.73 (4) $ & $782.65  \;^*$ & $0 \;^*$ & $773.51 (27)$  &  3.69  & $2.95 (8) $ & $1.44 (8) $ & $0.4379 (10)$ \\
 & CMD-2  & $0.93 (4)$ & $1.63 (5) $ & $782.65  \;^*$ & $0 \;^*$ & $774.40 (44)$  &  1.92  & $2.78 (10)$ & $1.28 (10)$ & $0.4375 (15)$ \\
 & BaBar  & $0.95 (1)$ & $2.09 (4) $ & $782.65  \;^*$ & $0 \;^*$ & $772.52 (10)$  &  1.12  & $3.57 (9) $ & $2.09 (11)$ & $0.4400 (4)$ \\
 & KLOE10 & $0.80 (1)$ & $1.69 (5) $ & $782.65  \;^*$ & $0 \;^*$ & $773.75 (17)$  &  1.06  & $2.89 (10)$ & $1.39 (10)$ & $0.4350 (7)$ \\
 & KLOE12 & $0.82 (2)$ & $1.51 (11)$ & $782.65  \;^*$ & $0 \;^*$ & $773.84 (27)$  &  1.21  & $2.57 (19)$ & $1.09 (17)$ & $0.4352 (10)$ \\
 & BESIII & $0.90 (4)$ & $1.87 (14)$ & $782.65  \;^*$ & $0 \;^*$ & $774.55 (41)$  &  0.74  & $3.19 (25)$ & $1.70 (26)$ & $0.4366 (14)$ \\
\midrule
Fit 2-$\rho$ 
 & SND    & $0.94 (2)$ & $1.81 (4) $ & $781.61 (10)$ & $0 \;^*$ & $773.96 (27)$  &  1.20  & $3.09 (9) $ & $1.60 (9) $ & $0.4381 (10)$ \\
 & CMD-2  & $0.97 (4)$ & $1.71 (6) $ & $782.30 (8) $ & $0 \;^*$ & $774.65 (44)$  &  1.29  & $2.92 (11)$ & $1.42 (11)$ & $0.4381 (15)$ \\
 & BaBar  & $0.95 (1)$ & $2.15 (4) $ & $781.78 (10)$ & $0 \;^*$ & $772.61 (10)$  &  0.83  & $3.67 (9) $ & $2.23 (11)$ & $0.4399 (4)$ \\
 & KLOE10 & $0.80 (1)$ & $1.71 (5) $ & $782.87 (14)$ & $0 \;^*$ & $773.77 (17)$  &  1.04  & $2.92 (10)$ & $1.41 (10)$ & $0.4350 (7)$ \\
 & KLOE12 & $0.82 (2)$ & $1.49 (11)$ & $782.32 (41)$ & $0 \;^*$ & $773.82 (27)$  &  1.22  & $2.54 (20)$ & $1.08 (17)$ & $0.4352 (10)$ \\
 & BESIII & $0.90 (4)$ & $1.92 (14)$ & $781.80 (43)$ & $0 \;^*$ & $774.65 (42)$  &  0.68  & $3.27 (25)$ & $1.80 (28)$ & $0.4365 (14)$ \\
\midrule
Fit 1-$\phi$ 
 & SND    & $1.02 (2)$ & $1.80 (4) $  & $782.65  \;^*$ & $11(1)$ & $773.6\;^*$  &  1.37  & $3.07 (8) $ & $1.57 (9) $  & $0.4404 (5)$ \\
 & CMD-2  & $1.06 (3)$ & $1.69 (6) $  & $782.65  \;^*$ & $5 (1)$ & $773.6\;^*$  &  1.38  & $2.89 (11)$ & $1.38 (10)$  & $0.4413 (6)$ \\
 & BaBar  & $0.96 (1)$ & $2.31 (4) $  & $782.65  \;^*$ & $10(1)$ & $773.6\;^*$  &  1.07  & $3.94 (9) $ & $2.58 (12)$  & $0.4391 (2)$ \\
 & KLOE10 & $0.81 (1)$ & $1.68 (5) $  & $782.65  \;^*$ & $-1(2)$ & $773.6\;^*$  &  1.07  & $2.86 (9) $ & $1.36 (9) $  & $0.4353 (3)$ \\
 & KLOE12 & $0.83 (1)$ & $1.46 (10)$  & $782.65  \;^*$ & $1 (4)$ & $773.6\;^*$  &  1.23  & $2.49 (18)$ & $1.03 (15)$  & $0.4357 (3)$ \\
 & BESIII & $0.93 (4)$ & $1.77 (13)$  & $782.65  \;^*$ & $5 (4)$ & $773.6\;^*$  &  0.81  & $3.01 (23)$ & $1.51 (23)$  & $0.4382 (9)$ \\
\midrule
Fit 2-$\phi$ 
 & SND    & $0.99 (2)$ & $1.80 (4) $  & $781.99 (20)$ & $5  (2)$ & $773.6\;^*$ &  1.13  & $3.07 (8) $ & $1.57 (9) $  & $0.4396 (5) $ \\
 & CMD-2  & $1.06 (3)$ & $1.69 (6) $  & $782.60 (22)$ & $4  (3)$ & $773.6\;^*$ &  1.43  & $2.89 (11)$ & $1.38 (11)$  & $0.4412 (7) $ \\
 & BaBar  & $0.95 (1)$ & $2.32 (4) $  & $782.27 (14)$ & $7  (1)$ & $773.6\;^*$ &  1.05  & $3.95 (9) $ & $2.60 (12)$  & $0.4390 (2) $ \\
 & KLOE10 & $0.81 (1)$ & $1.69 (5) $  & $782.96 (23)$ & $1  (3)$ & $773.6\;^*$ &  1.06  & $2.89 (10)$ & $1.38 (9) $  & $0.4354 (3) $ \\
 & KLOE12 & $0.83 (1)$ & $1.45 (10)$  & $781.92 (72)$ & $-4 (6)$ & $773.6\;^*$ &  1.23  & $2.47 (18)$ & $1.02 (15)$  & $0.4356 (3) $ \\
 & BESIII & $0.91 (4)$ & $1.78 (13)$  & $781.65 (74)$ & $-3 (7)$ & $773.6\;^*$ &  0.79  & $3.03 (23)$ & $1.53 (23)$  & $0.4376 (10)$ \\
\bottomrule
\end{tabular*}
\end{table*}
We observe that the fits work well in some, but not in all cases: the $p$-values characterizing the goodness of
the fits in particular to the SND and BaBar data are tiny.
In addition, not all results are consistent with each other. Of particular
interest for this work is the coupling $g_{\omega\pi\pi}$, extracted from each value of $\kappa_1$ via Eq.~\eqref{kappa1}. 
\begin{figure}[t!] 
\centering
\includegraphics*[width=\linewidth]{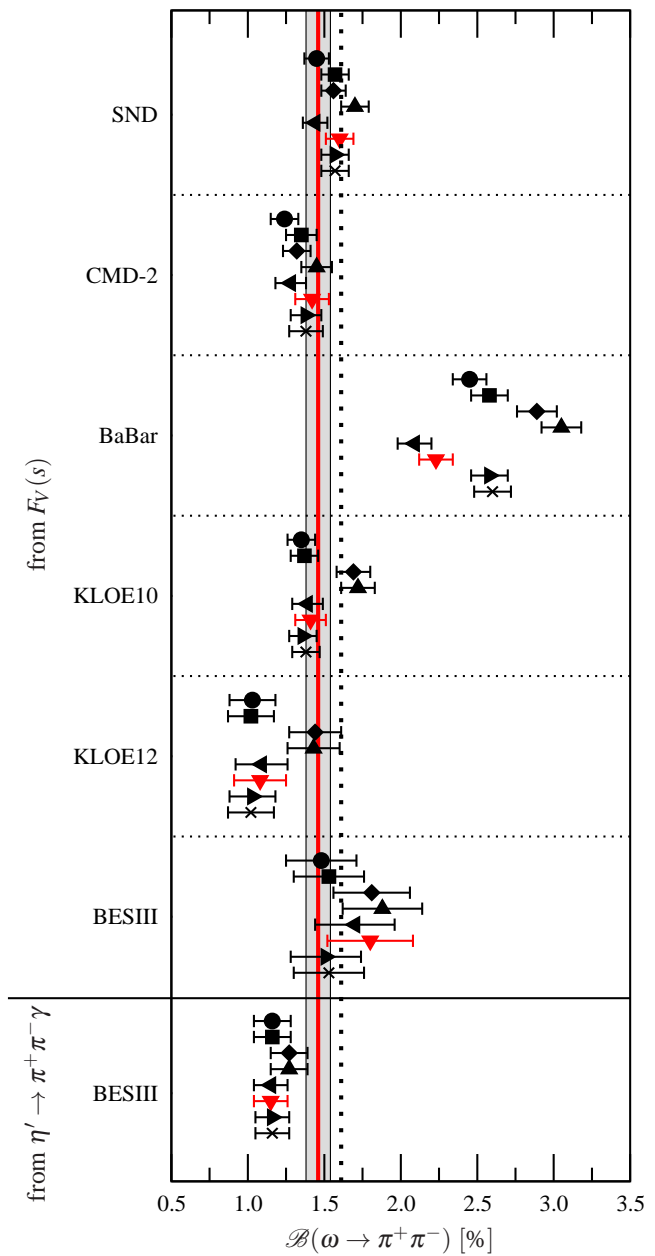}
\vspace*{0mm}
    \caption{Comparison of the values for the branching ratio for $\omega\to\pi^+\pi^-$ extracted from the various fits
    to the different data sets, where circles refer to Fit~1, squares to Fit~2, diamonds to Fit~I, the triangles-up to Fit~II,
    triangles-left to Fit~1-$\rho$,
    (red) triangles-down to Fit~2-$\rho$, triangles-right to Fit~1-$\phi$,  and crosses to Fit~2-$\phi$.  
    The red thick solid line denotes the average of the values, the gray band the corresponding uncertainty found from our preferred 
    analysis---Fit~2-$\rho$---omitting the contribution from BaBar. The average with the BaBar value included is shown
    as the perpendicular dotted line. Note that the values extracted from $\eta'\to\pi^+\pi^-\gamma$ refer to pseudo-data generated
    according to preliminary results.
    \label{fig:Bromegapipi}}
\end{figure}
Then using Eq.~\eqref{gammaompp} one can calculate the branching ratio for the transition $\omega\to\pi^+\pi^-$ from $g_{\omega\pi\pi}$. 
This quantity is also shown for all analyses in Table~\ref{tab:fitpvff} as well as in Fig.~\ref{fig:Bromegapipi}.
We find that most of the results appear consistent, however,
the branching ratio extracted from the BaBar analysis is significantly
larger than any of the other determinations. 

To better understand the reliability as well as
uncertainty of the extraction, we performed various additional fits. 
We account for the possibility of minor shifts in the experimental energy calibration, 
which may have consequences in particular in view of the narrowness of the $\omega$ signal.
Hence we repeated the fit described above, allowing the $\omega$ mass parameter to float. The
corresponding results are contained in Table~\ref{tab:fitpvff} as well as in Fig.~\ref{fig:Bromegapipi} as Fit~2.
We observe that the $\chi^2/\dof$ improves significantly in particular for the SND and BaBar data, which accordingly
are the only two sets for which the fitted $\omega$ mass deviates significantly from the PDG value
(taking into account both the fit errors and the uncertainty quoted by the PDG, $m_\omega = (782.65\pm0.12)\MeV$).
We note, however, that the extracted branching ratios $\Br(\omega\to\pi^+\pi^-)$ are stable throughout within one standard
deviation, even in the cases where the overall fit quality improves strongly.
We convinced ourselves that replacing the constant $\omega$ width by an energy-dependent
width as derived, e.g., in Ref.~\cite{Hoferichter:2014vra} changes the results negligibly.
\esp

In addition to the Madrid phase shifts~\cite{GarciaMartin:2011cn} used in most of the fits of our analysis, 
there is a second high-accuracy analysis of the $\pi\pi$ system available from the Bern group~\cite{Caprini:2011ky}. 
We thus also performed two fits using these phase shifts:
Fit~I is based on the $\omega$ mass as reported by the PDG, and
Fit~II allows for a floating $\omega$ mass. 
Overall, the resulting $\chi^2/\dof$ values tend to be a bit worse compared to the fits based on the Madrid phase shift; 
in particular, we cannot find acceptable $p$-values for fits to the BaBar data, not even with a floating $\omega$ mass.
The extracted $\omega\to\pi^+\pi^-$ couplings tend to be somewhat higher than in Fits~1 and 2; see also Fig.~\ref{fig:Bromegapipi}.
Varying the input phase around its central solution within the corresponding uncertainty band in a simplified, linearized
manner, we can slightly improve on the fit quality, but not by much; $g_{\omega\pi\pi}$ does not change beyond its error quoted
for the various Fits~I and II in Table~\ref{tab:fitpvff}.
This is most likely not the optimal way to utilize form factor data to fine-tune the Bern phase-shift
solution; a more sophisticated attempt to this end is currently under way~\cite{Colangelo:2017qdm}.

In principle, the pion vector form factor provides one of the most precise sources of information on the $\pi\pi$ $P$-wave
interactions, so one could turn the argument around and actually improve the precision of the phase shift $\delta_1(\spp)$ 
by adapting it to these data.  This has in fact already been done for the Madrid phase-shift 
analysis~\cite{DeTroconiz:2001rip,deTroconiz:2004yzs}, based on older form factor data.
Ref.~\cite{GarciaMartin:2011cn} provides an analytic parametrization for $\delta_1(\spp)$---cf.\ Eq.~(A7) of this
reference---that explicitly contains a mass parameter for the $\rho$-meson.
This parameter denotes the energy at which the phase shift passes through $\pi/2$ (and is therefore not to be confused
with the real part of the pole position of the $\rho$); 
its allowed range is quoted as $m_\rho = (773.6 \pm 0.9)\MeV$. 
In an attempt to optimize the phase shift ourselves in the fit to the pion form factor, 
we also allowed $m_\rho$ to float. 
The corresponding fit results appear in Table~\ref{tab:fitpvff} and Fig.~\ref{fig:Bromegapipi} 
as Fit~1-$\rho$ and Fit~2-$\rho$ for a fixed and a floating $\omega$
mass, respectively. 
Fit~2-$\rho$ finally is flexible enough to yield good fits with reasonable $p$-values for all six data sets.
It is interesting to observe that in all cases but for the fit to the BaBar data, the fits of the $\rho$-mass parameter 
overlap well within uncertainties with the range given by the analysis of Ref.~\cite{GarciaMartin:2011cn}. 

In the case of the BaBar data we found that the best fit is achieved when both the $\rho$ and the $\omega$ mass parameter are shifted downwards by
about $1\MeV$. 
This is in contrast to, e.g., the SND data, where the shift in $m_\omega$ is also large; however the one in $m_\rho$ is not (and tends to go
in the opposite direction).  
This might suggest that indeed some calibration problem is the origin of the incompatibility of the BaBar results with
the remaining data sets; such an explanation has been suggested before~\cite{Benayoun:2015gxa}.  
We could show, however, that at least the extracted value for $g_{\omega \pi\pi}$, the main focus of the present study,
is still rather insensitive to this (potential) issue: it changed only insignificantly
when we re-calibrated the BaBar data by a constant energy shift, adjusted such that the fit returns the central value
of the $\omega$ mass.
Finally, one might wonder whether the larger value of $g_{\omega \pi\pi}$ as extracted from the BaBar analysis is a consequence of the
higher energy resolution of that experiment. 
To test this hypothesis, we combined the BaBar bins in pairs, thus doubling the bin size, and redid the analysis.
This again led to an insignificant shift in the extracted value of $g_{\omega \pi\pi}$.

\begin{figure} 
 \centering
   \includegraphics*[width=\linewidth]{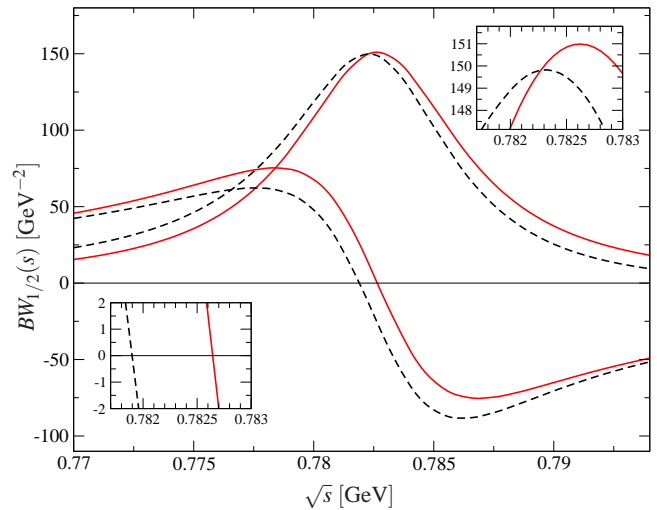}
   \caption{The line shapes of the amplitudes $BW_1(\spp)$ and $BW_2(\spp)$ defined in  Eqs.~\eqref{bws}
   are shown as (red) solid and (black) dashed lines. The peaked lines refer to the imaginary parts,
   while the ones with a zero refer to the real parts.
   The inserts magnify the regions of the zero-crossing of the real parts (bottom left) as well 
   as the maximum of the imaginary parts (top right). }\label{fig:breitw}
\end{figure}

\bsp
As discussed in Sect.~\ref{sec:FSI}, the fitting parameters used in our analysis in general,
and $\kappa_1$ in particular,
are necessarily real-valued as a consequence of 
unitarity. Contrary to this, in many experimental analyses a complex-valued coupling for $\omega\to\pi^+\pi^-$ is allowed. 
In order to demonstrate the stability and consistency of our results, we therefore redid Fits~1 and 2, however, now allowing for
a complex phase (sometimes called Orsay phase) attached to $g_{\omega\pi\pi}$. The results are reported in both Table~\ref{tab:fitpvff} as well as 
Fig.~\ref{fig:Bromegapipi} as Fit~1-$\phi$ and Fit~2-$\phi$. 
One observes that for the three newest data sets (KLOE10, KLOE12, and BESIII)
the fits returned phases consistent with zero. However, for the fits to the data by
SND, CMD-2, and BaBar in particular, Fit~1-$\phi$ shows phases 
that are nonzero by many standard deviations. In contrast, Fit~2-$\phi$, where the $\omega$ mass parameter was
allowed to float, yielded phases for SND and CMD-2 that are only marginally different from 0---the analysis of the BaBar
data requires a nonvanishing phase also in this case. The fact that the phases for the SND and CMD-2 fits become consistent with
zero once the $\omega$ mass is allowed to float is an illustration of the observation that Breit--Wigner parameters are
reaction-dependent: a phase in the coupling has a similar effect as a shift in the $\omega$ parameters. This is also illustrated
in Fig.~\ref{fig:breitw}, where we compare real and imaginary parts of two Breit--Wigner functions, namely
\begin{align}
BW_1 (\spp) &=\frac{1}{m_\omega^2-\spp-im_\omega \Gamma_{\omega}^{\rm tot}} ,  \nl
BW_2 (\spp) &=\frac{e^{i\, \phi}}{m_\omega^2-\spp-im_\omega \Gamma_{\omega}^{\rm tot}}  ,
\label{bws}
\end{align}
using the PDG values for $\omega$ mass and width for illustration, as well as $\phi=10^\circ$.
As Fig.~\ref{fig:breitw} demonstrates, 
introducing this phase in the coupling shifts the peak location of the imaginary part by $0.37\MeV$ to smaller
values of the energy, while the zero in the real part is shifted by $0.75\MeV$ in the same direction. Note that both
shifts are significantly larger than $0.12\MeV$, the uncertainty currently quoted for the $\omega$ mass
by the PDG. 
The isospin-violating effect that occurs in the pion vector form factor is (dominantly) sensitive to the 
real part of the Breit--Wigner amplitude, while reactions in which the $\omega$ is seen in the $3\pi$ channel
are largely sensitive to its imaginary part. 
The fact that the analysis of the BaBar data calls for a nonvanishing phase in the $\omega\pi\pi$ coupling even if the $\omega$
mass is allowed to float again points at some inconsistency of those data.

Finally, we also investigated the effect of an additional $\beta_V s^2$-term in Eq.~\eqref{eq:rspp}. 
We found that although this adds an additional free parameter to the analysis, the $\chi^2/\dof$ changed only 
marginally for all fits. In addition, the change in the value for $\kappa_1$ turned out to be entirely negligible 
compared to the quoted uncertainty.

We are now in the position to combine the results from the different experiments. 
The fits with the least bias are provided by Fit~2-$\rho$. A weighted average
of those results, omitting the result from the BaBar experiment, gives
\begin{equation}
\Br(\omega\to\pi^+\pi^-) = (1.46\pm 0.08) \times 10^{-2}  ,
\label{finalBr}
\end{equation}
where the uncertainty was scaled by a factor 1.5, applying the standard method of the 
PDG (described in detail in the introduction of the Review of Particle Physics~\cite{Olive:2016xmw}).
The result reported in Eq.~\eqref{finalBr} is consistent with the PDG average of $(1.49\pm 0.13)\%$~\cite{Olive:2016xmw},
however, with a somewhat reduced uncertainty.
We omit the BaBar results from the average on account of the following arguments that seem to indicate
an inconsistency within that data set, discussed in detail in this section:
\begin{enumerate}
\item the optimal $\omega$ mass is outside the range suggested by the PDG;
\item the optimal $\rho$ mass parameter in the $\pi\pi$ $P$-wave phase parametrization is outside the range
determined in Ref.~\cite{GarciaMartin:2011cn};
\item Fit 2-$\phi$ calls for a statistically significant nonvanishing complex phase of the coupling $g_{\omega\pi\pi}$, which is at odds with 
unitarity as long as the phase motion of the dominant (isospin-conserving) signal is under control, as it is
in our analysis;
\item the BaBar data set is the only one that does not seem to allow for an extraction of the branching
fraction $\Br(\omega\to\pi^+\pi^-)$ that is reasonably stable under the different fit variants, see Fig.~\ref{fig:Bromegapipi}.
\end{enumerate}
If we keep the BaBar data in the average, the branching ratio goes up to $(1.61\pm 0.15)\%$, with a scaling factor
larger than 3. In addition to the theoretical problems, this therefore also points 
at some inconsistency of the BaBar result with the other experiments.
\esp

\subsection{$\eta'\to\pi^+\pi^-\gamma$}
\bsp
While the large number of high-quality data sets on $e^+e^-\to\pi^+\pi^-$ clearly makes this a preferred reaction to extract
$\Br(\omega\to\pi^+\pi^-)$, it appears to be advisable to access the isospin-violating $\omega\to\pi^+\pi^-$ decay amplitude 
also from different reactions.  Besides aiming for a further improvement in the statistical precision of the determination
of this quantity, we may find further, systematically independent justification for our conclusion on the data selection
in the average, namely the omission of the BaBar results.
One future option could be the decay $\bar B_d^0\to J/\psi \pi^+ \pi^- $, where the mixing signal 
shows up very prominently~\cite{Daub:2015xja}.
However, the data presently available in this channel~\cite{Aaij:2014siy} are insufficient for a quantitative analysis.

An alternative is the very recent data on the radiative $\eta'$ decay
$\eta' \to \pi^+ \pi^- \gamma$ from BESIII. 
We have generated pseudo-data from the preliminary results presented in Ref.~\cite{Fang:2015vva},
where a model-independent fit of a functional form very similar to Eqs.~\eqref{eq:def_R+P} and \eqref{eq:ppspp} was used
(with, in view of the discussion in the previous subsection, mass and width of the $\omega$ fixed to their respective
PDG values).
\mbox{BESIII} has a data sample of about $9.7\times 10^5$ $\eta' \to \pi^+ \pi^- \gamma$ signal events in 100 energy bins,
with very low background (about 1\%) and a nearly flat acceptance; therefore, pseudo-data using $9.7\times 10^5$ events
should represent the statistical properties of the data set very well. 
We have performed an analogous series of eight fits as to the form factor data (with fixed and floating $m_\omega$,
Madrid and Bern phase input to the Omn\`es function, fitting $m_\rho$ inside the Madrid phase parametrization, and 
allowing for an Orsay phase $\phi$ multiplying the mixing term).
The main difference is that the polynomial $P(\spp)$ has a free normalization constant $A$ as well as a curvature term $\propto \beta$; 
see Eq.~\eqref{eq:ppspp}.
All fits were further constrained by the integrated partial width
$\Gamma (\eta' \to \pi^+\pi^- \gamma) = (0.0574 \pm 0.0028)\MeV$~\cite{Olive:2016xmw}.
Given that we are fitting pseudo-data, it is little surprising that $m_\omega$, $m_\rho$, 
and $\phi$ all come out consistently with their physical values in the cases where they are allowed 
to float. We mainly include these alternative
fits to illustrate the sensitivity of the data to these parameters.

\begin{figure} 
 \centering
   \includegraphics*[width=\linewidth]{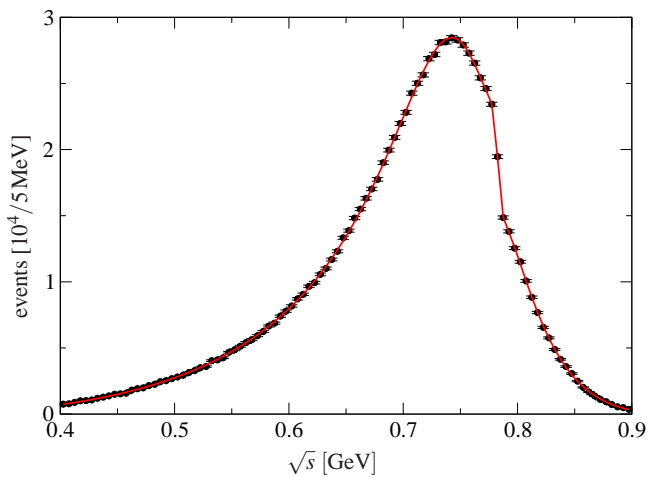}
   \caption{Best fit to the differential decay rate $\diff\Gamma(\eta'\to\pi^+\pi^-\gamma)/\diff \sqrt{\spp}$. 
            Pseudo-data generated according to preliminary BESIII results~\cite{Fang:2015vva}.}\label{fig:dcrosec}
\end{figure}
\begin{table*} 
\centering
\caption{Fit results for the BESIII~\cite{Fang:2015vva} pseudo-data of the $\eta' \to \pi^+ \pi^- \gamma$ decay spectrum. Fits~1--II and Fits 1-$\rho$--2-$\phi$ are analogous to what is described in Table~\ref{tab:fitpvff}.}
\label{tab:fitetappg}
\begin{tabular*}{\textwidth}{@{\extracolsep{\fill}}lllll llllll@{}}
\toprule
fits & $A$ & $\alpha$ &  $\beta$ & $\kappa_2 \times 10^3$ & $m_\omega$ & $\phi$ & $m_\rho$ & $\chi^2/\dof$ & $g_{\omega\pi\pi} \times 10^2$ & $\Br(\omega\to\pi^+\pi^-)$  \\
 & $[\GeV^{-3}]$ & $[\GeV^{-2}]$ &  $[\GeV^{-4}]$ & $[\GeV^{-1}]$ & $[\MeV]$ & $[{}^\circ]$ & $[\MeV]$ &  &  & $[\%]$  \\
\midrule
Fit 1 
 & $5.05 (13)$  & $0.99 (4)$  & $-0.55 (4)$  & $6.72 (24)$ & $782.65  \;^*$ & $0 \;^*$ & $773.6\;^*$  &  1.01  & $2.65 (13)$  & $1.16 (12)$  \\
Fit 2 
 & $5.05 (13)$  & $1.00 (4)$  & $-0.55 (4)$  & $6.72 (24)$ & $782.78 (14) $ & $0 \;^*$ & $773.6\;^*$  &  1.01  & $2.65 (13)$  & $1.16 (12)$  \\
Fit I 
 & $4.88 (13)$  & $1.18 (5)$  & $-0.82 (5)$  & $6.84 (24)$ & $782.65  \;^*$ & $0 \;^*$ & ---  &  1.32  & $2.69 (13)$  & $1.27 (12)$  \\
Fit II 
 & $4.88 (13)$  & $1.18 (5)$  & $-0.82 (5)$  & $6.84 (24)$ & $782.69 (14) $ & $0 \;^*$ & ---  &  1.34  & $2.69 (13)$  & $1.27 (12)$  \\
Fit 1-$\rho$
 & $5.08 (14)$  & $0.95 (7)$  & $-0.50 (8)$  & $6.69 (24)$ & $782.65  \;^*$ & $0 \;^*$ & $773.39 (29)$  &  1.02  & $2.63 (13)$  & $1.15 (11)$  \\
Fit 2-$\rho$ 
 & $5.08 (14)$  & $0.95 (7)$  & $-0.50 (8)$  & $6.68 (24)$ & $782.79 (15) $ & $0 \;^*$ & $773.36 (29)$  &  1.02  & $2.63 (13)$  & $1.15 (11)$  \\
Fit 1-$\phi$ 
 & $5.05 (13)$  & $1.00 (4)$  & $-0.56 (4)$  & $6.71 (24)$ & $782.65  \;^*$ & $-1(1)$ & $773.6\;^*$  &  1.01  & $2.64 (13)$  & $1.16 (11)$  \\
Fit 2-$\phi$ 
 & $5.05 (13)$  & $0.99 (4)$  & $-0.55 (4)$  & $6.72 (24)$ & $782.79 (23) $ & $ 0(2)$ & $773.6\;^*$  &  1.02  & $2.65 (13)$  & $1.16 (11)$  \\
\bottomrule
\end{tabular*}
\end{table*}

The optimal fit to these pseudo-data is shown in Fig.~\ref{fig:dcrosec}. 
The resulting fit parameters as well as the corresponding values for the minimal 
$\chi^2/\dof$ are displayed in Table~\ref{tab:fitetappg}. 
They confirm one major finding that was already firmly established for the closely related decay 
$\eta\to\pi^+\pi^-\gamma$~\cite{Stollenwerk:2011zz,Adlarson:2011xb,Babusci:2012ft,Kubis:2015sga}:
the parameter $\alpha$ is large, about an order of magnitude larger than the corresponding parameter $\alpha_V$
in the form factor fits.  
Here, however, the BESIII data for the first time demonstrate the necessity of the inclusion of the quadratic term $\propto \beta s^2$
with very high significance.  The leading left-hand-cut contribution provided by $a_2$-exchange gave an estimate 
of this parameter, $\beta = (-1.0\pm0.1)\GeV^{-4}$~\cite{Kubis:2015sga}, which yields the correct sign and order of magnitude, but 
is somewhat larger than what the new data suggest.  

In Table~\ref{tab:fitetappg} we also show the various values of $g_{\omega\pi\pi}$, extracted from $\kappa_2$ using Eq.~\eqref{kappa2},
as well as the results for $\Br(\omega\to\pi^+\pi^-)$,
which are also added at the bottom of Fig.~\ref{fig:Bromegapipi}.
Here, the variation of coupling constant and branching ratio is entirely negligible over the different fit variants.
Although we have only analyzed preliminary pseudo-data at present, 
the key message is that data of this quality are sufficient to provide an alternative
access to the isospin-violating decay $\omega\to\pi^+\pi^-$ with an accuracy comparable to that of form factor measurements. 
In addition, the 
experimental analysis currently available provides a clear preference for smaller values of $\Br(\omega\to\pi^+\pi^-)$, 
potentially even somewhat below the average cited in Eq.~\eqref{finalBr}, and definitely in contradiction to the large
numbers found based on the BaBar form factor data.
\esp

\section{The pion charge radius}\label{sec:radius}

On the basis of the present analysis we are now also in the position to extract an improved value for the
pion vector radius.
It is understood as the square root of the mean squared radius $\langle r_V^2\rangle$,
which in turn is defined by the polynomial expansion of the form factor $F_V(\spp)$ around $\spp=0$, 
\begin{equation}
F_V(\spp) = 1 + \frac{1}{6}\langle r_V^2\rangle \spp + \Order(\spp^2) .
\end{equation}
Within the formalism introduced above it may be written as
\begin{equation}
\langle r_V^2\rangle = \frac{6}{\pi} \int_{4m_\pi^2}^\infty\diff x\frac{\delta_1(x)}{x^2} 
+ 6 \bigg (\alpha_V + \frac{\kappa_1}{m_\omega^2} \bigg) , \label{eq:radius}
\end{equation}
where the first term stems from the expansion of the Omn\`es function, and we have neglected tiny corrections scaling with 
the $\omega$ width in the isospin-breaking contribution (which is very small to begin with). 
The ratio of two Omn\`es functions calculated employing two moderately differing high-energy continuations of the phase shifts 
has a polynomial form at low energies. 
Since the parameter $\alpha_V$ is determined via a fit to data it therefore implicitly also depends
on the high-energy behavior assumed for the phase shifts. However, the pion radius is necessarily independent thereon.\footnote{In fact,
we have also performed fits with a pion form factor phase (instead of the elastic scattering phase shift) 
as input to the Omn\`es function, including effects of the 
$\rho(1450)$ and $\rho(1700)$ resonances; see Ref.~\cite{Schneider:2012ez} for details.  This reduces the parameter $\alpha_V$ almost
to zero, however, the radii come out consistent with the present analysis in line with the reasoning given.}

\begin{figure} 
 \centering
   \includegraphics*[width=\linewidth]{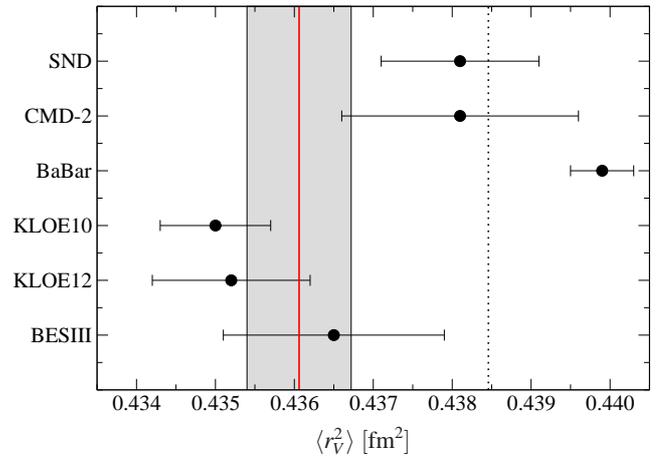}
    \caption{Comparison of the values for the pion charge radius extracted from the analysis of
     the different vector form factor data sets. Only the results of our preferred Fit~2-$\rho$ are shown.
     The red thick solid line denotes the average of the values, the gray band the corresponding uncertainty found 
     omitting the contribution from BaBar. The average with the BaBar value included is shown
     as the perpendicular dotted line.
     \label{fig:radius}}
\end{figure}
For the study of the radius we again only use the results of our preferred fit, namely Fit~2-$\rho$. 
In order to control the effect of possible correlations between the fitted value of the $\rho$ mass parameter $m_\rho$
and the parameter $\alpha_V$ on the radius, we performed two additional fits to each data set, where 
we fixed $m_\rho$ to its corresponding minimal and maximal value allowed by  Fit~2-$\rho$. The uncertainty 
of the radius is then determined for each experiment from the largest spread in the radii allowed in those fits.
The results are shown in Table~\ref{tab:fitpvff} and Fig.~\ref{fig:radius}.

Finally also for the study of the radius we investigated the effect of an additional $\beta_V s^2$-term in Eq.~\eqref{eq:rspp} and observed that the uncertainty in $\alpha$ increased by a factor of 4--10 (combined with an almost unchanged $\chi^2/\dof$), depending on the data set, while the values of $\beta_V$ turned out consistent with 0 within 1--2$\sigma$.  The only exception is once more the BaBar data set that calls for a nonzero value for $\beta_V$ by about $4.5\sigma$. In addition, the central values determined for the radius including the $\beta_V$-term are in most cases consistent with those from the original fit within $1\sigma$. We therefore do not quote the results of these additional analyses in detail.

\begin{figure}
\centering
\includegraphics*[width=0.9\linewidth]{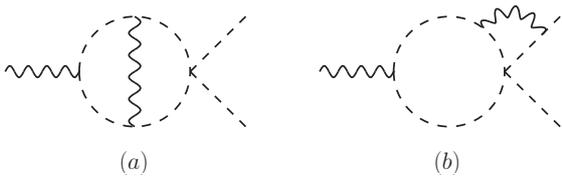}
\caption{Examples for Feynman diagrams in chiral perturbation theory, leading to nonuniversal radiative corrections.  
Dashed lines refer to charged pions, wiggly lines to photons.}\label{fig:radcorr}
\end{figure}

\bsp
The statistical uncertainty of the squared radius extracted from the fit turns out be of the order of 
0.2\%. At this level of accuracy one needs to worry also about effects from radiative corrections.
The generic size of one-loop corrections in the squared radius is given by $6\alpha_{\rm em}/(4\pi m_\pi^2)$
(see e.g.\ Ref.~\cite{Kubis:1999db}), which is much larger than the statistical uncertainty.
However, these are all universal in the sense that they can be formulated in terms of an overall multiplicative
factor, calculable in scalar QED, and were already removed in the experimental analyses when extracting the form factor
from the cross sections.
On the other hand, there are additional $s$-dependent terms induced by nonuniversal terms---see,
e.g., Fig.~\ref{fig:radcorr}---, which in the framework of chiral perturbation theory appear 
at two-loop order.  These may affect the extrapolation of the form factor from the physical, timelike region
($\spp > 4m_\pi^2$) to $\spp=0$, where the radius is defined; note, for instance, that the diagram 
Fig.~\ref{fig:radcorr}(\textit{a}) contains a logarithmic singularity at threshold; see e.g.\ the detailed
discussion in Ref.~\cite{Bissegger:2008ff}.\footnote{We point out that in general, it is not possible to isolate
purely hadronic quantities in the presence of electromagnetic interactions~\cite{Gasser:2003hk}.} 
We therefore estimate the possible effect of an additional $s$-dependence on the squared radius, induced by 
nonuniversal radiative corrections, by those obtained from chiral perturbation theory two-loop diagrams such 
as those in Fig.~\ref{fig:radcorr}, which generically scale as $6\alpha_{\rm em}/(\pi\Lambda^2)$, 
where $\Lambda \approx 1\GeV$ denotes the characteristic scale for the chiral expansion.

Averaging the fit results to the individual experiments, omitting again the result from BaBar for the reasons
discussed in Sect.~\ref{sec:formfactor}, we find
\begin{equation}
\langle r_V^2\rangle = (0.4361\pm 0.0007\pm 0.0005) \fm^2 ,
\end{equation}
where the first error denotes the statistical uncertainty given by the fit---it includes a scale factor of 1.5 determined according to the procedure proposed by the PDG---and the second one the uncertainty by possible residual radiative corrections estimated above. Our result is consistent with the allowed parameter range for the squared radius between $0.42\fm^2$ and $0.44\fm^2$ derived on very general grounds in Ref.~\cite{Ananthanarayan:2013dpa}. 
This translates into
\begin{equation}
\sqrt{\langle r_V^2\rangle} = (0.6603\pm 0.0005\pm 0.0004) \fm 
\end{equation}
for the radius. This value is to 
be compared to the current PDG average $(0.672\pm 0.008)\fm$. Both values agree
within $2\sigma$, however, our number has a significantly reduced uncertainty.
It is also interesting to remark that if one keeps only those values in the average
quoted in the Review of Particle Physics that were extracted from $e\pi\to e\pi$ (which basically
means omitting values extracted from $eN\to e\pi N$ that might contain some additional model
dependence not included in the uncertainty~\cite{Liesenfeld:1999mv}), the average
drops to $(0.663\pm 0.006)\fm$, fully in line with the value quoted above, however, with
a significantly larger uncertainty. 
Had we kept the BaBar result, the radius would have shifted to
$\langle r_V^2\rangle = (0.4385\pm 0.0009\pm 0.0005)\fm^2$, which translates to ${\langle r_V^2\rangle}^{1/2} = (0.6622\pm 0.0007\pm 0.0004) \fm$, however,
here again a scaling factor of 3.3 was necessary for the uncertainty, once more pointing at an inconsistency of the 
BaBar data compared to the others. This inconsistency is also quite apparent in Fig.~\ref{fig:radius}.
\esp

\section{Conclusion}\label{sec:summary}

\bsp
Exploiting the universality of final-state interactions by means of dispersion theory as well as the analytic structure
of the pion vector form factor and the amplitude for $\eta'\to\pi^+\pi^- \gamma$, we extracted information on
the branching fraction $\Br(\omega\to \pi^+\pi^-)$ and the pion charge radius. Our analysis shows
that the BaBar form factor data~\cite{Aubert:2009ad} are inconsistent with the other analyses
as well as with theoretical constraints in various respects, but in particular concerning the
$\omega\to\pi^+\pi^-$ coupling strength. 
It should be noted that other groups came to similar conclusions; see, e.g., 
Refs.~\cite{Ananthanarayan:2012tt,Benayoun:2012wc}. 
We therefore do not include the BaBar form factor data in our final averages.
\esp

Based on recent data from SND, CMD-2, KLOE, and BESIII, we found $\Br(\omega\to\pi^+\pi^-) = (1.46\pm 0.08) \times 10^{-2}$
and $\sqrt{\langle r_V^2\rangle} = 0.6603(5)(4)\fm$, where the first uncertainty is statistical based on the 
data fits, and the second estimates the size of nonuniversal radiative corrections. 
Both values are consistent with those currently reported
by the PDG~\cite{Olive:2016xmw}, however, with reduced uncertainties. Only one of the experiments included
in our study has been included in the PDG average for $\Br(\omega\to\pi^+\pi^-)$ so far, and none for the pion charge radius.

We have finally pointed out that high-quality data on $\eta'\to\pi^+\pi^-\gamma$ will allow one to further improve on
the value for $\Br(\omega\to\pi^+\pi^-)$, and cross-check the consistency of the different pion form factor data sets.
Final data for this decay can be expected in the very near future from both the CLAS~\cite{Kunkel:2016wrp} and the BESIII~\cite{Fang:2015vva} 
collaborations.

\begin{acknowledgements}
\bsp
CWX thanks A.~Nogga and J.~L.~Wynen for useful discussions and kind help. 
Furthermore, we are grateful to G.~Colangelo and P.~Stoffer for providing us with the pion--pion phase-shift
solution of Ref.~\cite{Caprini:2011ky}.
We thank G.~Venanzoni, S.~E.~M\"uller, T.~Teubner, and A.~Keshavarzi for useful communication.
This research is supported in part by the DFG and the NSFC through funds provided to the 
Sino--German CRC~110 ``Symmetries and the Emergence of Structure in QCD'' 
(NSFC Grant No.\ 11621131001, DFG Grant No.\ TRR110),
and by the National Science Foundation under Grant No.\ NSF PHY-1125915.
\esp
\end{acknowledgements}

\newpage

\section*{Erratum to: The branching ratio \boldmath{$\omega\to \pi^+\pi^-$} revisited}

\setcounter{figure}{0}
\setcounter{table}{0}
\setcounter{equation}{0}
\setcounter{footnote}{0}

\bsp
In the published paper~\cite{E-Hanhart:2016pcd}, a data set
for the pion vector form factor from the BaBar experiment~\cite{E-Aubert:2009ad} was used 
that was not consistent with the ones employed for the other experiments, since it did not include the contributions from the vacuum polarization,
while it contained those of the final-state radiation. The use of the official form factor data
changes some of the results, as we report in this erratum.
\esp

\begin{figure}[t!] 
\centering
\includegraphics*[width=\linewidth]{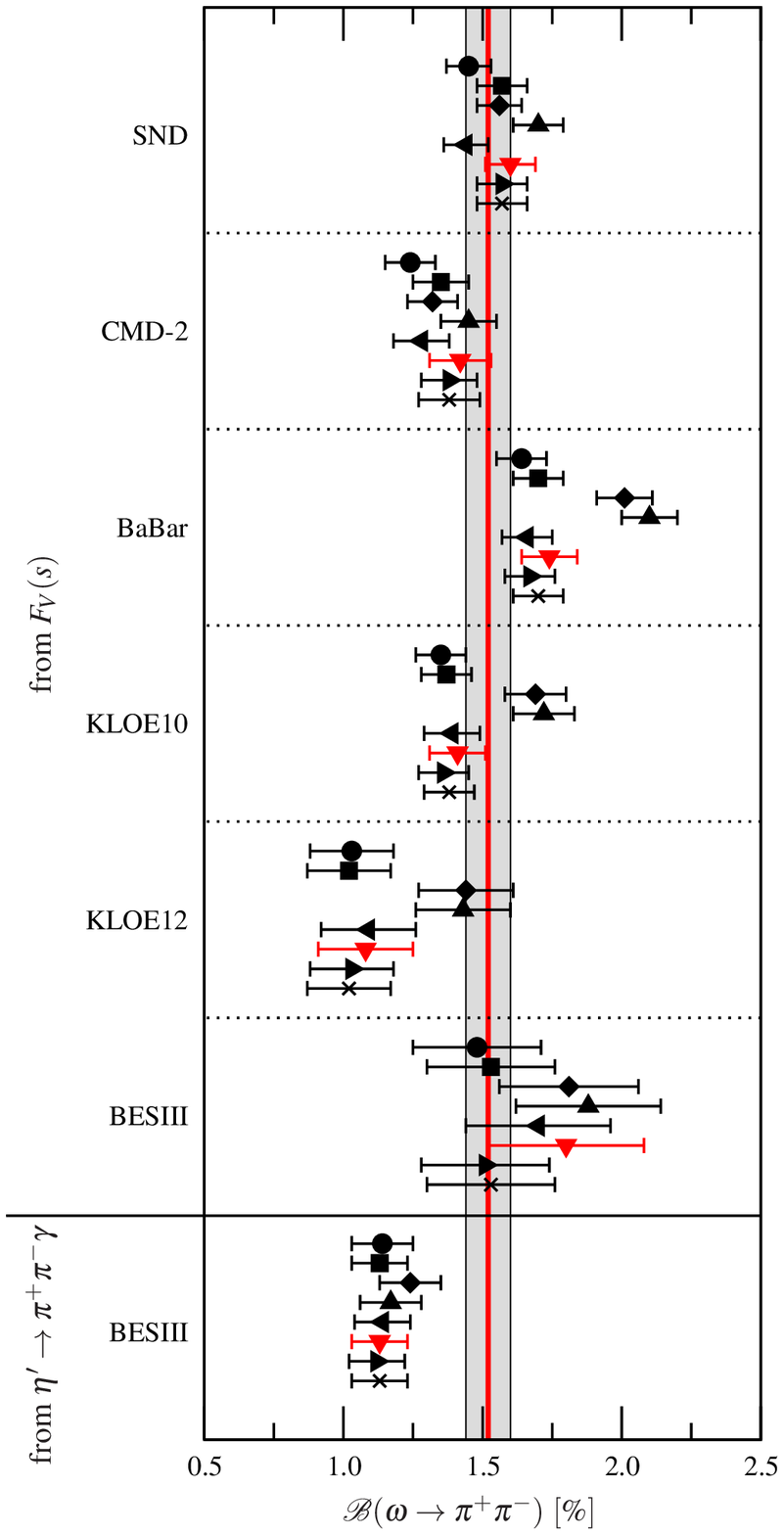}
\vspace*{0mm}
    \caption{Comparison of the values for the branching ratio for $\omega\to\pi^+\pi^-$ extracted from the various fits
    to the different data sets, where circles refer to Fit~1, squares to Fit~2, diamonds to Fit~I, the triangles-up to Fit~II,
    triangles-left to Fit~1-$\rho$,
    (red) triangles-down to Fit~2-$\rho$, triangles-right to Fit~1-$\phi$,  and crosses to Fit~2-$\phi$.  
    The red thick solid line denotes the average of the values, the gray band the corresponding uncertainty found from our preferred 
    analysis---Fit~2-$\rho$---including all data sets. 
    Note that the values extracted from $\eta'\to\pi^+\pi^-\gamma$ refer to pseudo-data generated
    according to preliminary results.
    \label{E-fig:Bromegapipi}}
\end{figure}

In Table~\ref{E-table} we report the results of the various fits to the corrected data set as well as the
quantities derived from the fit parameters like the $\omega\to \pi\pi$ branching fraction and the 
square of the pion radius. The results for the branching fractions extracted from the different fits
are compared to those extracted from the other data sets in Fig.~\ref{E-fig:Bromegapipi}.

\begin{table*}
\renewcommand{\arraystretch}{1.2}
\caption{
Fit results for the pion vector form factor using the corrected BaBar data~\cite{E-Aubert:2009ad,E-Lees:2012cj}. \label{E-table}}
\begin{tabular*}{\textwidth}{@{\extracolsep{\fill}}llll lllll l@{}}
\toprule
Fits & $\alpha_V \times 10$ & $\kappa_1 \times 10^3$ & $\quad m_\omega$ & $\;\phi$ & $\quad m_\rho$ & $\chi^2/\dof$ & $g_{\omega\pi\pi} \times 10^2$ & $\Br(\omega\to\pi\pi)$ & $\quad \langle r_V^2 \rangle$ \\
 &  $[\GeV^{-2}]$ &  & $[\MeV]$ & $[\;^\circ\;]$ & $[\MeV]$ &  &  & $\quad\%$ & $\quad [\text{fm}^2]$ \\
\midrule
Fit 1 
   & $1.08 (1)$ & $1.84 (4) $ & $782.65 \;^*$ & $0 \;^*$ & $773.6\;^*$  &  0.90  & $3.14 (8) $ & $1.64 (9)$ & $0.4418 (2)$ \\
Fit 2 
  & $1.08 (1)$ & $1.87 (4) $ & $781.86 (12)$ & $0 \;^*$ & $773.6\;^*$  &  0.72  & $3.19 (8) $ & $1.70 (9)$ & $0.4419 (2)$ \\
Fit I
  & $0.67 (1)$ & $1.99 (4) $ & $782.65 \;^*$ & $0 \;^*$ & ---  &  1.23  & $3.40 (8) $ & $2.01 (10)$  & $0.4381 (2)$ \\
Fit II 
  & $0.67 (1)$ & $2.03 (4) $ & $781.79 (11)$ & $0 \;^*$ & ---  &  1.00  & $3.46 (8) $ & $2.10 (10)$  & $0.4381 (2)$ \\
Fit 1-$\rho$ 
  & $1.08 (1)$ & $1.85 (4) $ & $782.65 \;^*$ & $0 \;^*$ & $773.69 (10)$  &  0.90  & $3.16 (9) $ & $1.66 (9)$ & $0.4417 (4)$ \\
Fit 2-$\rho$ 
  & $1.08 (1)$ & $1.90 (4) $ & $781.85 (11)$ & $0 \;^*$ & $773.76 (10)$  &  0.72  & $3.23 (9) $ & $1.74(10)$ & $0.4416 (4)$ \\
Fit 1-$\phi$ 
  & $1.10 (1)$ & $1.86 (4) $ & $782.65 \;^*$ & $6 (1)$ & $773.6\;^*$ &  0.80  & $3.17 (8) $ & $1.67 (9)$  & $0.4422 (2) $ \\
Fit 2-$\phi$ 
  & $1.08 (1)$ & $1.87 (4) $ & $781.88 (17)$ & $0 (2)$ & $773.6\;^*$ &  0.73  & $3.19 (8) $ & $1.70 (9)$  & $0.4419 (2) $ \\
\bottomrule
\end{tabular*}
\end{table*}

It is obvious from Fig.~\ref{E-fig:Bromegapipi} that with the corrected data set for the BaBar experiment, the fluctuations of
the results for the $\omega\to \pi\pi$ branching ratio derived from
the different fit strategies are reduced significantly to a level compatible with those of the other experiments.
Furthermore, the branching ratio extracted is now much more in line with the other experiments than
before. Note that the difference between the branching ratio $(1.74\pm 0.10)\%$, quoted in Table~\ref{E-table},  
and the  branching ratio  $(1.44\pm 0.11)\%$ given in Ref.~\cite{E-Davier:2013rla} can be partially
traced back to our using the $\omega$ width as an input (fixed from other experiments, in particular 
$e^+e^-\to3\pi$), while it was varied in the analyses of Refs.~\cite{E-Aubert:2009ad,E-Davier:2013rla}.

Throughout this erratum we stick to the strategy already chosen in Ref.~\cite{E-Hanhart:2016pcd},
namely that the final results are derived from weighted averages of the findings of Fit~2-$\rho$.

We then find for the $\omega\to\pi\pi$ branching ratio with the corrected BaBar data included
\begin{equation}
\Br(\omega\to\pi^+\pi^-) = (1.52\pm 0.08) \times 10^{-2}  ,
\label{E-finalBrwB}
\end{equation}
where, according to the prescription of the Particle Data Group (cf.\ the introduction of Ref.~\cite{E-Patrignani:2016xqp}), 
a scaling factor of $1.8$ had to be applied to the uncertainty.

This is to be compared to the value extracted with the BaBar data omitted
\begin{equation}
\Br(\omega\to\pi^+\pi^-) = (1.46\pm 0.08) \times 10^{-2}  ,
\label{E-finalBr}
\end{equation}
where a scale factor of $1.5$ was applied. The latter value 
was already reported in Ref.~\cite{E-Hanhart:2016pcd}.
Clearly, once the correct BaBar data set is employed in the analysis, the $\omega\to\pi\pi$ 
branching ratio comes out statistically consistent with the other values. Moreover, 
the mass parameters of both the $\omega$ and the $\rho$ are found to be reasonably consistent with
those of the other extractions, and the fit does not call for a complex phase of the
coupling $g_{\omega\pi\pi}$, in line with unitarity. From this point of view there is no reason anymore to prefer
the value given in Eq.~\eqref{E-finalBr} to the one of Eq.~\eqref{E-finalBrwB}.

The situation is somewhat different for the pion charge radius. The central value for the radius derived
from the BaBar data now even more strongly deviates from the values derived from the other experiments:
our preferred fit (Fit 2-$\rho$) results in (cf.\ Table~\ref{E-table})
\begin{equation}
\langle r_V^2\rangle = (0.4416 \pm 0.0004 \pm 0.0005) \fm^2 ,
\label{E-radiusB}
\end{equation}
where the first error denotes the statistical uncertainty given by the fit and the second one the uncertainty by possible 
residual radiative corrections estimated in Ref.~\cite{E-Hanhart:2016pcd}. 
This is to be compared to the value extracted from an average over the other experiments
 with the BaBar data omitted~\cite{E-Hanhart:2016pcd}
\begin{equation}
\langle r_V^2\rangle = (0.4361\pm 0.0007\pm 0.0005) \fm^2 ,
\label{E-radiuswoB}
\end{equation}
where a scale factor of 1.5 was included in the uncertainty as detailed in Ref.~\cite{E-Hanhart:2016pcd}.\footnote{We 
would like to stress that this number cannot be directly compared to the radius reported in Ref.~\cite{E-Ananthanarayan:2017efc}, 
$\langle r_V^2\rangle = (0.4320 \pm 0.0041)\fm^2$,
since the latter number was deduced from form factors with vacuum polarization removed prior to the analysis.}
We emphasize that the first uncertainties quoted in Eqs.~\eqref{E-radiusB} and~\eqref{E-radiuswoB} are
based solely on the statistical errors derived from the fits performed with a given fixed set of
parameters for the $\pi\pi$ phase shifts~\cite{E-GarciaMartin:2011cn}. 
In particular the ranges for those parameters also quoted in Ref.~\cite{E-GarciaMartin:2011cn} were
not considered. Preliminary studies trying to include these uncertainties in the analysis
using Bayesian statistics indicate that the full uncertainty might be dominated by the 
systematics resulting from this procedure. We therefore neither perform
a combined fit for the radii nor quote a final result at this stage. The aforementioned more advanced
studies indicate that the uncertainty for the $\omega\to \pi\pi$ branching fraction changes
only marginally.

Note that the central value for the square of the pion radius derived from the BaBar data using the parametrization 
of Ref.~\cite{E-Lees:2012cj} reads
\begin{equation}
\langle r_V^2\rangle = (0.433 + i \ 0.004) \fm^2 .
\end{equation}
Ref.~\cite{E-Malaescu:thesis} quotes the value $(0.4319\pm 0.0016)\fm^2$ for the real part of $\langle r_V^2\rangle$.
The difference in the central value may be understood from different rounding procedures.
In this reference no number is
given for the imaginary part, but we expect its uncertainty to be of comparable size.
Thus the real part of the pion radius derived from the parameterization employed in Ref.~\cite{E-Lees:2012cj} 
is even lower than
the number reported in Eq.~\eqref{E-radiuswoB}, however, it comes with a nonvanishing imaginary part in conflict with 
general principles, which demonstrates that this parametrization
is inapt to derive a reliable value of the charge radius---while a single 
Gounaris--Sakurai term is consistent with unitarity and analyticity, a sum of those
terms with complex coefficients as used in Ref.~\cite{E-Lees:2012cj} violates both.

\begin{acknowledgements}
\bsp
We are grateful to M.~Davier and B.~Malaescu for providing us with the form factor data derived from 
Ref.~\cite{E-Aubert:2009ad}; see also Ref.~\cite{E-Lees:2012cj}. 
We would like to thank B.~Ananthanarayan, M.~Davier, S.~Eidelman, M.~Hofe\-richter, F.~Ignatov, I.~Logashenko, 
B.~Malaescu, and G.~Venanzoni for useful discussions
and extensive e-mail communication.  
\esp
\end{acknowledgements}


\end{document}